\documentclass[prl,aps,twocolumn,amsmath,amssymb,nofootinbib,tighten,floatfix]{revtex4}
\usepackage{graphicx}
\usepackage{wasysym}
\usepackage{subfigure}

\newcommand{\beq}{\begin{equation}}
\newcommand{\eeq}{\end{equation}}
\newcommand{\bea}{\begin{eqnarray}}
\newcommand{\eea}{\end{eqnarray}}

\newcommand{\bei}{\begin{itemize}}
\newcommand{\eei}{\end{itemize}}

\def\tit#1#2#3#4#5{{#1}{\bf #2}, #3 (#4)}

\def\rmp{Rev.\ Mod.\ Phys.\ }

\def\prl{Phys.\ Rev.\ Lett.\ }

\def\prb{Phys.\ Rev.\ B\ }

\def\mrb{Mat.\ Res.\ Bull.\ }

\def\jetp{Sov.\ Phys.\ JETP\ }

\begin{document}

\title{Irrational charge from topological order}

\author{R. Moessner$^\dag$ and S. L. Sondhi$^\ddag$}
\affiliation{$^\dag$Max-Planck-Institut f\"ur Physik komplexer Systeme, 01187
  Dresden, Germany\\
$^\ddag$Department of Physics, Princeton University, NJ 08544, USA}


\begin{abstract}
Topological or deconfined phases of matter exhibit emergent gauge fields and
quasiparticles that carry a corresponding gauge charge. In systems
with an intrinsic conserved $U(1)$ charge, such as all electronic systems where
the Coulombic charge plays this role, these quasiparticles are also
characterized by their intrinsic charge. We show that one can take advantage of
the topological order fairly generally to produce periodic Hamiltonians which
endow the quasiparticles with continuously variable, generically irrational,
intrinsic charges. Examples include various topologically ordered lattice
models, the three dimensional RVB liquid on bipartite lattices as well as water
and spin ice. By contrast, the gauge charges of the quasiparticles retain their
quantized values.
\end{abstract}
\maketitle

\noindent
{\bf Introduction:}
The phenomenon of quantum number fractionalization~\cite{rr-review} is among
the most striking collective phenomena in
condensed matter physics: the low-energy excitations of a many-body system can
exhibit quantum numbers  which cannot be obtained by simple addition or
subtraction of those of the high-energy degrees of freedom. The best-known
example is charge fractionalization: excitations can carry fractional charge,
e.g.\  $e/3$ in the case of Laughlin quasiparticles, even though we are quite
confident that the underlying degrees of freedom are electrons with quantized
elementary charge $e$.

In one dimension fractionalization is associated with soliton formation, but in
the known examples in higher dimensions it goes along with phases of matter
that exhibit emergent gauge fields~\cite{wenbook}. Such phases are variously
called fractionalized, deconfined or topological; the last term is the one we
shall use in this paper even though strictly it refers only to phases where the
emergent gauge field is described by a purely topological
action~\cite{nayaketalrmp}. As a consequence, the low energy excitations are
classified not only by their fractionalized intrinsic quantum numbers but also
by their emergent gauge charge(s) and/or fluxes under the emergent gauge
field(s).

In the case of the Laughlin quasiparticles, this distinction is not crucial as
they carry a gauge flux and an intrinsic electric charge that are slaved to
each other by a constraint~\cite{zhk}. However, as we show here, in
general this distinction {\it is} important:  While the emergent gauge quantum
numbers of a state
can effectively be considered to be fixed~\cite{hastingsUT}, it is possible to
tune the intrinsic quantum numbers continuously by perturbing the Hamiltonian.

Specifically, we consider systems that are characterized by a microscopically
conserved $U(1)$ charge which we call the intrinsic charge. When the Coulomb
interaction is explicitly included, this charge is, as usual, the source of an
electric field; in other cases it is simply proportional to a conserved
particle number. We show that the intrinsic charge of the quasiparticles in our
examples can take on continuous, generally irrational, values. This generalizes
the one dimensional irrational charge found by Brazovskii~\cite{brazovskii} and
by Rice and Mele~\cite{ricemele} to higher dimensions and {\it ex post facto}
reinterprets their result as establishing the independence of the intrinsic
charge from the topological charge characterizing solitons in their problem. [We
also note the stimulating work by Chamon and collaborators \cite{chamon1,chamon2}
who have generalized the field theoretic treatment~\cite{jackiw} of solitonic
fractionalization and the Brazovskii--Rice-Mele effect to two dimensions. Interestingly,
they find logarithmically confined excitations with continuously variable charge
which re-rationalizes upon inclusion of a gauge field to deconfine them.]

In the balance of the paper we describe this phenomenon in more detail in three
different contexts. First, we introduce a variant of the well studied quantum
dimer model on hypercubic lattices which allows a controlled computation of the
irrational charge. Next, we consider the short ranged RVB liquid phase on
hypercubic lattices in $d \geq 3$ and show that its excitations can be made to
carry irrational intrinsic charges in the presence of an (inevitable)
electron-phonon coupling. Finally, we review the corresponding physics in water
and spin ice, which exhibit a classical limit of topological order, and provide
experimentally realized instances of this phenomenon.

\noindent
{\bf Topologically ordered lattice models:}
We first present a simple lattice model for which an explicit calculation is
possible. This model describes bosons with charge $q$, created by
$b_{i\alpha}^\dagger$, living on the links of a hypercubic lattice in $d \ge
3$. The links are labeled by the sites they join, the Roman/Greek letters
referring to the A/B sublattices of these bipartite lattices, with Hamiltonian
\bea
H=H_U+H_{QDM}+H_s\ ,
\label{eq:irratbosonH}
\eea
\bea
H_U = U \sum_{i} \sum_{\alpha,\beta \in nn(i)} (b_{i\alpha}^\dagger b_{i\alpha}) (b_{i\beta}^\dagger b_{i\beta}) \ ,
\eea

\bea
H_{QDM}=
-&t& \sum_{p} \sum_{\stackrel{\{i,j,\alpha,\beta\}}{\in p}} b^\dagger_{i \beta} b^\dagger_{j \alpha}
b_{j \beta} b_{i \alpha} + h. c.  \\
+&v& \sum_{p} \sum_{\stackrel{\{i,j,\alpha,\beta\}}{\in p}} (b^\dagger_{i \alpha} b_{i \alpha})
(b^\dagger_{j \beta} b_{j \beta}) + \{ i\leftrightarrow j\} \nonumber
\label{eq:HQDM}
\eea
\beq
H_s=-s \sum_{i} \sum_{\alpha,\beta \in nn(i)} b^\dagger_{i\alpha} b_{i\beta} +
h.c. \ .
\eeq
The sum on $i$ runs over all A sublattice sites, $nn(i)$ refers to the
B sublattice neighbors of site $i$, and $p$ indexes the plaquettes of the
lattice. The boson density per site is fixed to be 1/2. Note that the
boson number is conserved and defines the intrinsic charge in this problem.

We now consider the terms in the Hamiltonian in order of their strength, $U\gg
v,t\gg s$, initially ignoring $H_s$.

\begin{figure}[htp]
  \begin{center}
    \subfigure[]{\label{fig:edge-a}\includegraphics[width=3.5cm,angle=0]{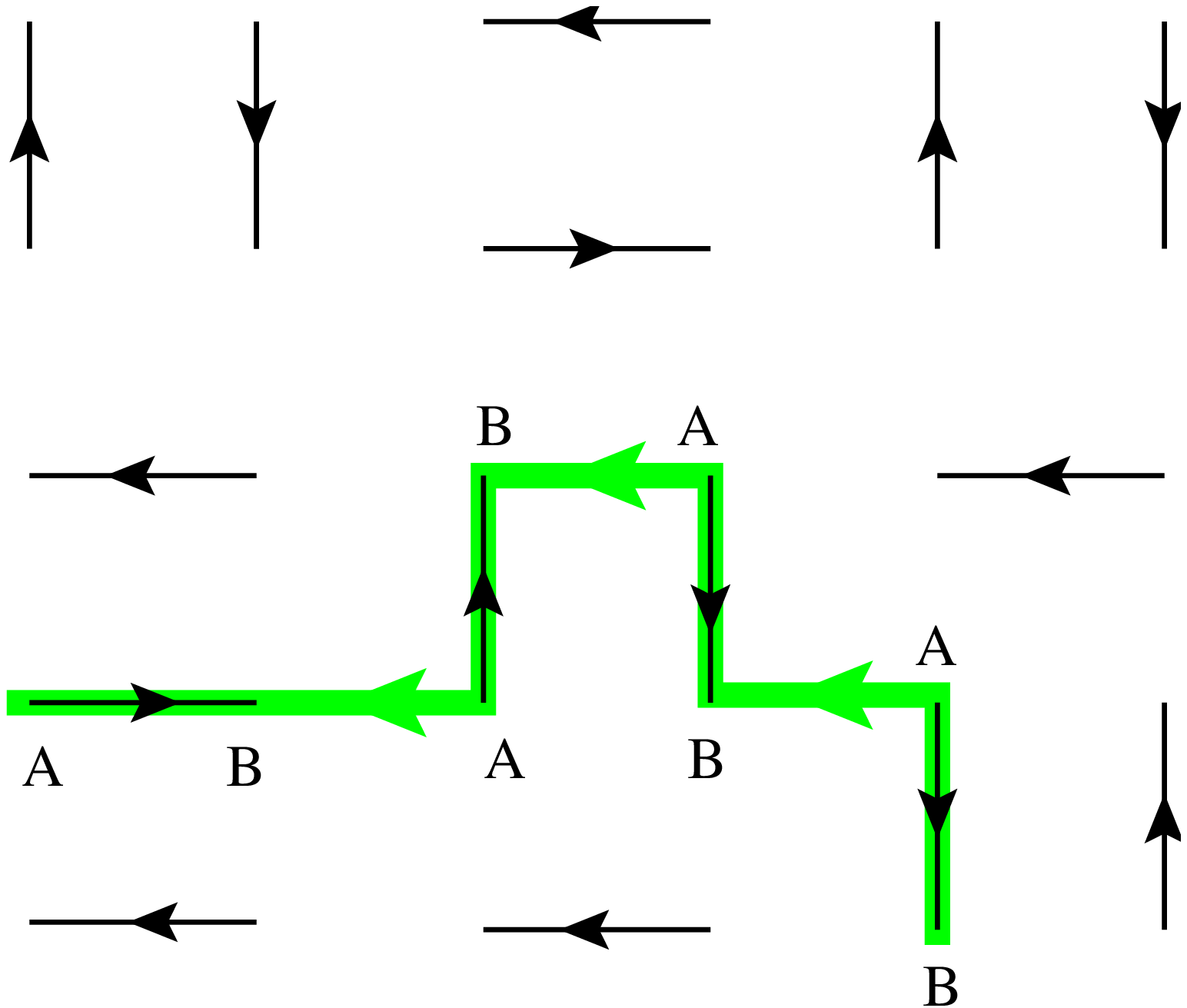}}\hfill
    \subfigure[]{\label{fig:edge-b}\includegraphics[width=3.5cm,angle=0]{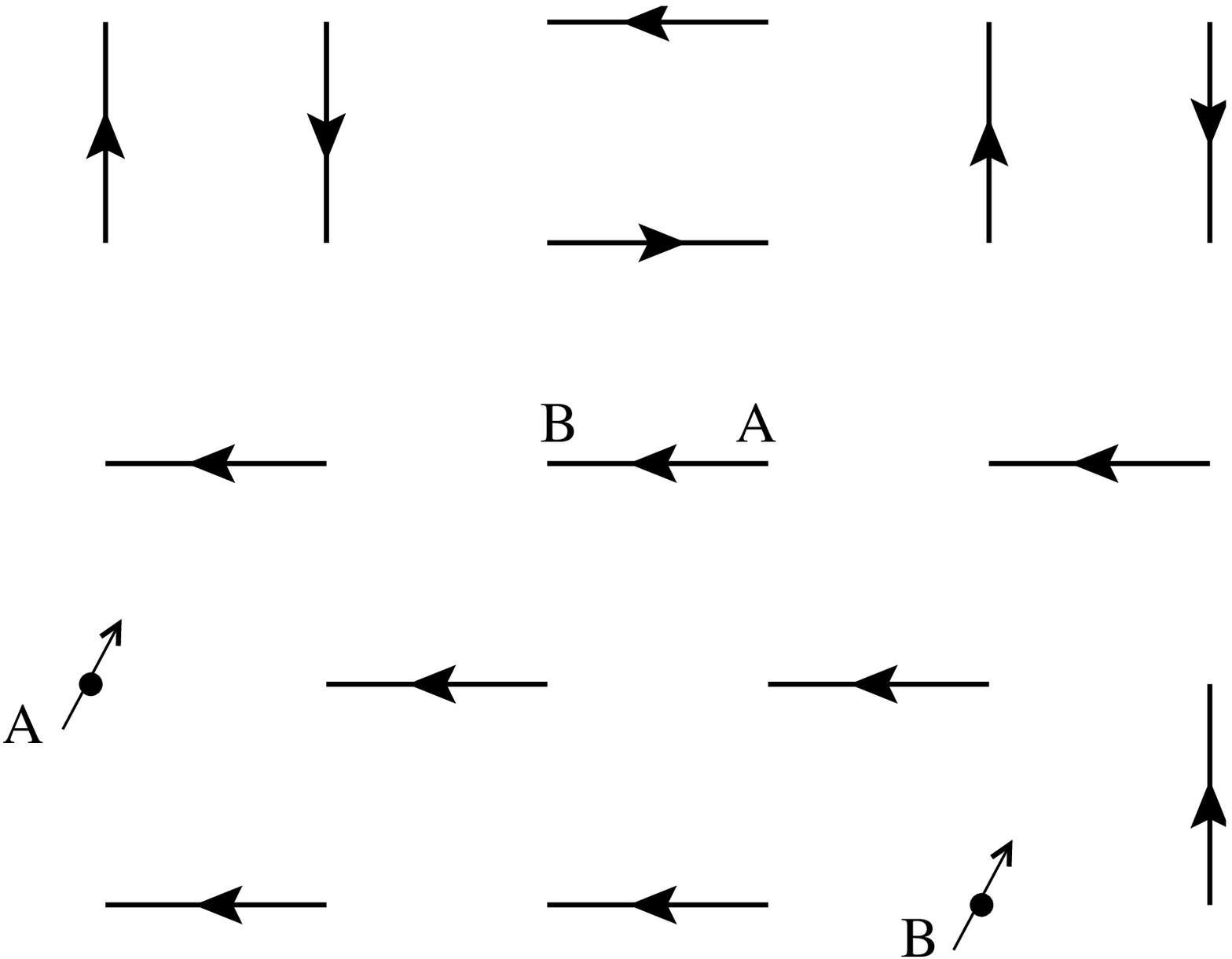}}
  \end{center}\vspace{-1cm}\mbox{}
  \caption{(a) A dimer covering representing either bosons living on the bonds
    or singlets formed between spins on the adjacent sites. The arrows indicate
    the orientation of the induced dipole moments on the dimers. The
    modification of the state along the (fat green) oriented string of dipoles
    generates the dimer configuration with two monomers/spinons shown in
    (b). We have indicated the sublattices where appropriate. We are displaying
    a two-dimensional cartoon, even though many of our results hold in $d\geq3$
    only. }
      \label{irratfig1}\vspace{-0.7cm}\mbox{}
\end{figure}

In this limit, $H_U$ singles out a low energy manifold of boson configurations
with exactly one boson occupying the bonds emanating from a given site---these
are in precise correspondence with hard core dimer coverings of the lattice,
Fig~\ref{irratfig1}. This manifold is separated from all excited states by a
gap; in our ordering of energy scales the leading order dynamics takes place
entirely within it. Indeed, the action of $H_{QDM}$ within this dimer manifold
is exactly that of the quantum dimer model which is known to exhibit a
deconfined $U(1)$ ``Coulomb'' phase for $v\alt t$ on the hypercubic lattice in
$d\geq3$~\cite{ms3drvb,henleyreview}.

The gauge field that is deconfined in this phase can be traced to the structure
of the dimer states as follows. Let us assign an oriented lattice flux to each
link~\cite{ms3drvb}:
\bea
F_{i \alpha}=b_{i\alpha}^\dagger b_{i\alpha}-\frac{1}{2d}=-F_{\alpha i} .
\label{eq:flux}
\eea
Each dimer configuration then has a vanishing lattice divergence ${\bf
  \nabla\cdot F}=0$ at each site of the lattice. The solution of this
sourceless ``Gauss's law'' constraint, ${\bf F} = \nabla \times {\bf A}$
uncovers the gauge field at issue. The low energy physics of the dimer manifold
is that of Maxwell theory with a gapless mode corresponding to the
photon~\cite{ms3drvb}.

In addition, there are point-like excitations just above the gap of order $U$,
violating Gauss's law, which are thus charged under $\bf A$. Specifically,
these sites are associated with monomers and trimers, i.e.\ they have no or two
dimers emanating from them. Crucially, there are {\it two} species of each
carrying opposite gauge charge, one for each sublattice. Monomers on sublattice
A/B (and trimers on sublattice B/A) act as sinks/sources of the ``electric''
flux $\bf F$. These particles interact much as their analogs in fundamental
electrodynamics.

Finally, let us identify the the intrinsic charge of the monomers and trimers
at this stage of the analysis. As a pair of monomers/trimers is created by
removing/adding one boson, they are seen to carry intrinsic charges $\mp
q/2$. While we have derived the gauge and intrinsic charges in a leading order
analysis in the large $U$ limit, these are exact results at large $U$ for the
combination $H_U + H_{QDM}$.

Now let us turn to the last term $H_s$, which breaks the sublattice symmetry
by introducing boson hopping only between pairs of links meeting in the $A$
sublattice. A characteristic property of topologically ordered systems is that
any small perturbation leaves the low energy structure qualitatively
intact. Thus quite generally we expect a sufficiently small value of $s$ to
lead to another Hamiltonian with a gapless photon and monomers/trimers with
the gauge charges identified above~\cite{hastingsUT}. For our  model, the
situation is particularly robust, as  $H_s$  has no matrix elements within the
dimer manifold. It is only at order $s^2/U$ that we get a renormalization of
$t$, innocuous at small $s/U$ as it moves the system to a nearby point in the
parameter space of the dimer model. Higher order corrections generate other
terms but the general conclusion will remain.

However, there is another effect of turning on $H_s$---it leads to a
redistribution of charge in the perturbatively modified dimer states. This
modifies the intrinsic charge of the gauge charged excitations. The easiest way
to compute the change in intrinsic charge is to observe that a given dimer in
each configuration acquires a dipole moment of magnitude $P = {1 \over 2} q a d
(s/U)^2$ ($a$ is the eventually irrelevant lattice constant). This moment is
oriented from sublattice A to B, and is computed with respect to its nominal
position at the center of a bond. The creation of a pair of monomers/trimers
then involves the insertion of a suitable string of such dipoles oriented head
to tail (see Fig.~\ref{irratfig1} for an example). Such a string exhibits a
pair of endpoint charges of magnitude $P/a$. Adding this correction to the
starting values we find that monomers/trimers now carry the intrinsic charges
\bea
Q_m/q = -{1 \over 2}  \pm  {d \over 2} ({s \over U})^2; \ \ \ 
Q_t/q = +{1 \over 2}  \mp  {d \over 2} ({s \over U})^2 \ ,
\eea
up to corrections of $O(s/U)^4$ on sublattice A/B. The reader can check that a
direct computation of the coarse grained intrinsic charge agrees with these
values.

The ingredients needed above to decouple gauge and intrinsic charges --
topological order and sublattice specific quasiparticles -- are also present in
various other models with topological phases to which our analysis can be
applied {\it mutatis mutandis}. Examples of these are the bosonic models
of~\cite{senthilmot,motrunich} which exhibit a conserved boson number in
addition to emergent $U(1)$ gauge fields coupled to Higgs scalars of various
charge. Specifically, our construction works readily for the  $Z_3 \times Z_3$
Higgs phase on a $d=2$ triangular lattice of~\cite{motrunich} and for the
Coulomb phase in $d \ge 3$ for the charge $3$ and higher generalizations of the
models presented in~\cite{senthilmot}.  Specifically, what is required in
these cases is the introduction of a staggered potential that breaks the
symmetry between sublattices. Finally our construction also generalizes to
quantum Hall states in the Tao-Thouless limit~\cite{HaldRez}.

This is also a good place to note that there are cases when our construction
does {\it not} work. These include the charge 2 Higgs condensed phase
of~\cite{senthilmot} and the deconfined phase in the triangular lattice quantum
dimer model~\cite{MStrirvb} both of which exhibit $Z_2$ topological order and
an absence of sublattice specific quasiparticles. The coarse grained intrinsic
charge in such cases cannot be altered by breaking sublattice symmetries.

\noindent
{\bf RVB on bipartite lattices:} The quantum dimer model was constructed to
capture the essential physics of the short ranged resonating valence bond
(SR-RVB) state proposed by Anderson~\cite{anderson} as a description of a
disordered, non-magnetic ground state of a Mott insulator -- the prototypical
spin liquid. As such,  our above analysis suggests that the phenomenon of
charge irrationalization should extend to SR-RVB physics in its electronic
setting. This is indeed true but the details work out somewhat differently, as
we describe next.

The largest energy scale in the Mott insulator is the charge gap $\Delta_c$
stemming from the Coulomb repulsion. Below it the dynamical degrees of freedom
in the singly occupied manifold are the spins of the electrons. In an SR-RVB
insulator, there is a second scale $\Delta_{\rm RVB}<\Delta_c$, which
characterizes the formation of local singlets, with each spin forming a singlet
with exactly one neighbor. At lower energies the basic degrees of freedom of
the SR-RVB models are thus again hard core dimers on bonds, but now they encode
the formation of a singlet bond between the two spins at the ends of the dimer,
as opposed to the boson living on the bond in our model above. A concomitant
key difference is that the conserved intrinsic charge is not the number of
dimers on bonds but instead the number of electrons on sites.

The physics in the valence bond/dimer manifold is again well captured by the
processes contained in $H_{QDM}$ (\ref{eq:HQDM}) and the states and spectra are
isomorphic~\cite{fn-rvbdetails}. Thus on hypercubic lattices in $d \ge 3$, the
low energy physics of the SR-RVB state is also described by an emergent $U(1)$
gauge field with a Maxwell action.

The gauge charged excitations are spinons which are the analogs of the monomers
above: spins which are not involved in a singlet bond.  Again they come in two
sublattice flavors with opposite gauge charges and {\it prima facie} they carry
spin $S_s=1/2$ and intrinsic charge $Q_s=0$ as the distribution of electronic
charge appears undisturbed by their formation.

However, that last statement ceases to be true when we add two ingredients to
this discussion. First, we follow our prescription above and explicitly break
the symmetry between the sublattices---this time via a weak staggered potential
which favors electronic occupancy of the $A$
sublattice~\cite{fn-poly}. Secondly, note that electron hopping between
neighboring sites is increased if they are connected by a spin singlet, because
both of the Pauli principle forbidding hopping between aligned spins, and due
to the electron-lattice coupling. These two ingredients combine so that each
dimer now involves not only a spin singlet but also a nonzero excess
probability $\delta$ of finding both electrons on its sublattice $A$ rather
than $B$ end. Thus dimers carry an electric dipole moment $P =2 e \delta
a$ oriented from sublattice B to A. As before, the result is to endow spinons
created by the insertions of strings of dipoles with charges $\pm Q_s= e
\delta$, which, being continuously tunable by means of the sublattice potential
and/or electron-lattice coupling, are in general again irrational. We note that
upon acquiring charge, the spinons contribute to the optical conductivity in
the Mott gap.

Finally, we would be remiss if we did not remind the reader that the SR-RVB
liquid~\cite{MStrirvb} on the non-bipartite triangular lattice, is free of
irrationality following our comments at the end of the last section.

\noindent
{\bf Water and spin ice:} The above constructions establish a point of
principle but they do not currently apply directly to an experiment. To
remedy this, we turn to a pair of materials one of which should be
familiar even to string theorists among our readers. We find that
deconfined irrational electric charge exists even in the absence of any
explicit sublattice symmetry breaking. This we trace to a classical limit of
$U(1)$ topological order.

The starting point of our analysis once again consists of identifying a
solenoidal vector field which can be done in either the cubic or hexagonal ice
phases I$_c$ or I$_h$~\cite{henleyreview}. In either phase, the structure of
ice can be described as a tetrahedrally coordinated network of oxygen ions with
single protons placed asymmetrically between the pair of oxygens they bond. The
ice rules state that exactly two protons sit close to each oxygen (thereby
forming an ``H$_2$O molecule''). Assigning a flux $\pm1$ to the links with a
close (distant) proton, this defines the solenoidal field we require. Note that the emergent ``electric'' flux on a bond comes automatically with an actual
electric dipole moment and thus we need not take recourse to the
sublattice constructions used earlier.

Bjerrum defects are the most common violation of the ice rules carrying a gauge
charge. They involve taking an ``H$_2$O molecule'' and rotating it so that one
of its bonds has no protons sitting on it, and another two  (one near and one
far). A simple point-charge model gives an irrational charge
of~\cite{nagleunit}
\bea
Q_{\rm Bjerrum}=\pm\sqrt{3}\mu/a\approx\pm 0.38 e\ ,
\eea
where $\mu$ is the dipole moment of the ``H$_2$O molecule'', $a$ the distance
between neighbouring oxygens, and $e$ the electronic charge.

Closely related to this are the ionic defects, a pair of which carrying
opposite gauge and electric charge can be created by finding an ``electric''
flux tube/dipole string and reversing it. This creates two excitations located
at the ends which carry gauge charge $\pm 2$ and electric charge
\bea
Q_{\rm ion}=\pm (|e|-|Q_{\rm Bjerrum}|) \ .
\eea
With $\mu$ an accident of local chemistry, and as the oxygen separation is
tunable by the application of pressure, it is clear that $Q_{\rm Bjerrum, ion}$
are continuously tunable as well.

In this discussion we appear to have dispensed with topological order
altogether but this is misleading. We need our defects to be deconfined. This
can be accomplished most robustly by picking a quantum Hamiltonian acting
on the ice ground state manifold that is in a topological
phase~\cite{fradkinneto}. In actual ice, instead, there is (to good accuracy) a
classical Hamiltonian -- that leaves ice rule compliant configurations
energetically degenerate -- which {\it also} exhibits a purely static emergent
gauge field that does not confine defects charged under it~\cite{fn-classconf}.

Thus ice also supports deconfined irrationally charged excitations on account
of (classical) topological order.

While the full context discussed in this paper is new, the fact that defects in
water ice carry irrational electric charge has been long
recognized~\cite{nagleunit}.  In chemistry, irrational charge is of course not
an unusual occurrence, as ions can `share' an electron in a chemical bond in a
non-quantized way. However, turning this charge into a deconfined quasiparticle
is not  trivial. The fact that common water ice pulls off this magic trick is
remarkable.

This story takes another fascinating twist in the case of spin ice,  a magnetic
analogue of water ice. The statistical mechanics of water ice is locally  that
of spin ice. The basic electric degrees of freedom, however, are replaced by
magnetic ones: Ising magnetic reside on the midpoints of the links of a diamond
lattice of bond length $a$, whose local easy axes are the bond directions.

This analogy led to the recent realization~\cite{cmsmonopoles} that there are
excitations in spin ice that carry irrational {\em magnetic} charge
$Q_{\rm monopole}=\pm2\mu/a\approx\pm q_D/8000$, where $q_D$ is the size of
Dirac's monopole. These magnetic monopoles interact Coulombically and are
sources of the magnetic field $H$ (rather than the magnetic flux density
$B$). In the absence of microscopic monopoles, magnetic charge arises from the
fractionalization of the dipoles.

\noindent
{\bf In closing:} The two central messages of this paper are first, that gauge
and intrinsic charges in topologically ordered systems are logically distinct
and second, that this distinction permits the intrinsic charge to take on a
continuous set of, generically, irrational values. We have presented examples
in $d=2$ and $d=3$ which involve $Z_3$ and $U(1)$ gauge fields in quantum and
classical settings. We expect that such examples could be multiplied for other
instances of topological order with gapped matter, including cases with
non Abelian gauge fields, and also for quantum numbers other than the intrinsic charge.

\noindent
{\bf Acknowledgements:}
We are very grateful to Claudio Castelnovo for collaboration on closely related
work, in particular Ref.~\cite{cmsmonopoles}; to Claudio Chamon for
introducing us to his and his colleagues's work on the quest for higher-dimensional irrational
charge~\cite{chamon1}; to Steve Kivelson for continuing support and
encouragement; and to all of them for useful discussions.


\begin{thebibliography}{99}
\bibitem{fn-poly}
{This is borrowed from the $d=1$ case of polyacetylene, where this happens naturally if the polymer chain consists of alternating species of atoms.}
\bibitem{fn-classconf} A lifting of the degeneracy of the ice configurations
  typically leads to the selection of an ordered state at $T=0$ in which the
  irrationally charged defects are confined. In practice, deconfinement
  therefore obtains in the temperature window above the scale of any
  degeneracy-lifting term, and below the melting transition of ice.
\end{thebibliography}
\end{document}